# Applications of quantum dots to the treatment of retinitis related disorders

*R. Dragomir*

**ABSTRACT**: *Retinitis is an inherited disease that affects roughly 1 in 4000 people worldwide. In this paper I suggest four possible applications of quantum dot technology in the treatment of retinitis related disorders such aș retinitis pigmentosa. The causes of this disease are not yet clear but it is accepted that necrotic rupture of the membrane and cone cell enlargement together with rod apoptosis are the main mechanisms of retinal cell degeneration.*

Yearly, about 1.5 million people in the world lose their sight from a disorder of the retina named retinitis pigmentosa (RP). The problem that this paper aims to solve is treating a genetic disorder, RP with the aid of quantum dot technology. A review of current concepts in the treatment for this disorder can be found in the medical literature [1], and available treatments range from gene therapy, retinal transplantation, vitamin therapy to stem cells. Recently it was found [2] that cone cells die by necrosis in RP while cone cells die by another mechanism that mainly involves membrane rupture and cell enlargement, but the exact cause of RP is yet to be elucidated. Genetic mutations of pigments of photoreceptor cells (PC) were also shown to produce certain forms of RP.

The retina of the human eye contains millions of PC that transform light into electro-chemical signals that are processed by the brain to produce the image we finally perceive. For example, the human eye contains about 130 million rods and about 7 million cones. Rod cells mediate black-and-white vision while cone cells mediate color vision. It is known that in RP the loss of color vision due to cone cell death is the most debilitating part of this disease.

I suggest here some applications of quantum dots (QD) in the treatment of RP.

These solutions have not been yet proposed in the literature.

**Quantum dot retinal implants**

Having briefly introduced the anatomy of the anterior part of the retina we will review the technology of semiconductor quantum dots. QDs are geometrical structures of semiconductor crystals. A subclass of QDs are the optically active quantum dots (OAQDs), representing only those QDs that can interact with light quanta (photons). A photon creates an exciton i.e. an electron-hole pair, where by hole we understand a vacancy in the valence band of the semiconductor. Then an applied field along the QD may disintegrate the exciton via electron and hole quantum tunneling,respectively into electron and hole reservoirs. This in turn gives rise to a measurable photocurrent. Briefly, light can be converted to current using OAQD (see Fig. 1).

It is important to mention that OAQDs can be made biocompatible and used in vivo. For example, quantum dots that have a spherical shape and that are made of CdSe are nowadays encapsulated in ZnS for keeping the toxic $Cd^{2+}$ ions from being freed into the human body. I mention that the compound is not toxic at concentrations of maximum 10nM(i.e. 0.01mg/ml). Another method used to prevent the oxidative release of $Cd^{2+}$ ions into the human body is the polymer coating of the QDs [3]. The resulting compound can thus be made biocompatible at even greater solution concentrations: 1 μM (i.e. 1mg/ml). The previously fabricated QDs were still found to be unstable at room temperature.

The cell experiments in [4] revealed that chitosan (CS) coated QDs (CS-ZnS and CS-ZnS:$Mn^{2+}$) showed low cytotoxicity and good biocompatibility making them convenient for in vivo use.

After the coating of the quantum dots made of ZnS or ZnS: $Mn^{2+}$ with a hidrophilic coating such as chitosan I propose that the QDs to be further conjugated with streptavidin molecule that attach to the biotinylated anti-rhodopsin.

In [5] shows that by covalently linking phenothiazine (PTZ) through phenyldithiocarbamate (PTC) to a QD made of CdS we can allow outside tunneling of holes on a subpicosecond timescale. In [6] it was reported that ZnS QDs can also be linked to PTZ in order to make possible the hole tunneling. Using molecular electrons acceptors that are physiosorbed on the QD surface, electrons can tunnel outside into the surrounding solution on the single-picosecond timescale.

The solution of QDs will target the damaged PC cells in the retina if they are inserted in the retina. The insertion is not harmful to the body if less than $2x10^9$ of QDs are injected[7]. The compunds arrives at the place of the ruptured membrane of the PC and insert themselves into the plasmalema of the PC because rhodopsin pigments are to be found there. Then the bias of chemical potentials between the extracellular fluid and the intracellular fluid transforms the impinging light quanta (coming from the aqueuos vitreous) into a photocurrent. This current gives rise to a graded action potential along the PC thus restoring the function of the previously unhealthy cell.

I assume that the photocurrent induced by the exciton generating light should be of the range of pA while the bias between extracellular fluid and intracellular fluid is expected to be of about several mV. About $10^3$ photons impinging per second on a PC will generate tens pA of current across the membrane which should be sufficient to give rise to a graded action potențial.

I mention that there are three types of anti-bodies for the three types of rhodopsin found in a cone cell. They facilitate the binding to the pigment with the right color. In order to make the process complete QDs must be grown in three sizes, depending on the detected wavelength of the absorbed photon.

One remark is that one has to ensure itself that a QD compound from the colloidal solution has a low enough molecular mass to permit the crossing of the blood-retina-barrier (BRB) of the solution. This should not present an impediment în the case of diabetic retinitis where BRB is highly permeable.

The suggested treatment în this section involves the restoring of the dead PC. Colloidal OAQDs made of ZnS, coated with hidrophylic chains of chitosans and then conjugated with streptavidin, biotin and anti-bodies should be able to target the rhodopsin pigments in the retinal cone cells. Once bound to the specific rhodopsin molecule in the PC the OAQD should start to function in the transport regime in which photons are converted to electrical signals that regenerate the lost action potential needed for the functioning of the retina. We hope that the solution presented in this section would present better prospects for rehabilitation of a patient who suffers from RP.

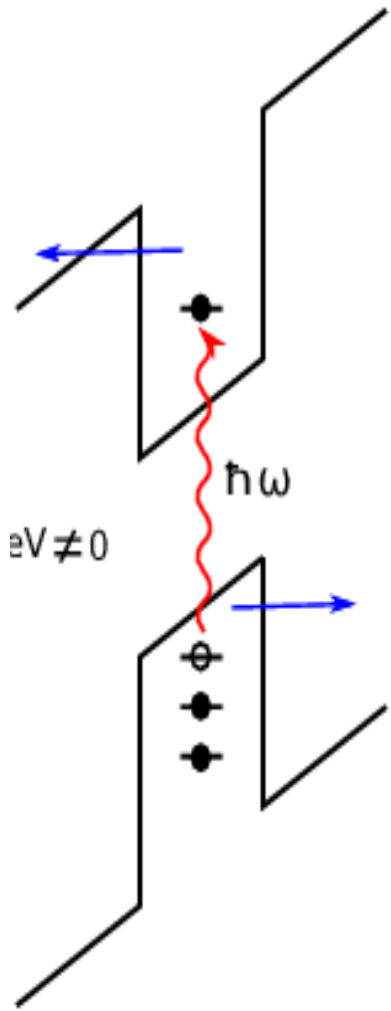

Fig. 1: Exciton formation and disintegration in quantum dots followed by photocurrent generation.

**ATP synthesis using quantum dots**

As we have seen in the previous sections PC are of two types: rods and cones. Rods mediate the color vision at high light intensity while rods mediate the black and white vision at low light intensity.

In RP, PC exhibit cellular enlargement, along with necrotic changes such as cellular swelling and mitochondrial rupture. Mitochondrial rupture leads to the depletion of ATP in the cone cell. Since ATP drives the pumps that generate a gradient of ions across the PC. Mitochondrial rupture leads to the absence of an action potential along the PC. ATP is formed by the binding ADP to the phosphate (Pi), and at its hydrolization energy is released.

I mention here that OAQD can have tunable size and shape therefore I will consider two conical OAQDs (see Fig 2a). Light entering the eye on these two dots should be able to generate excitons (pairs of electron-hole). I will also consider that the QDs are made of a semiconductor material in which the hole's effective mass is larger than the electron's effective mass (e.g. CdTe, CdSe). Because the energy of a particle that is confined in a potential well of nanometric dimension is inversely proportional to the product between the mass of that particle and the square of the length of the well, and because the hole's effective mass is larger than the electron's mass, the probability density of finding a hole in the QD will distribute to the vertex of the conical quantum dot. The electron's probability density will distribute towards the base of the conical QD. Therefore by shining of light on these conical shaped QD we will be to generate electric dipoles. Two such electric dipoles will attract each other as described in Fig. 2 b) because the lowest energy configuration is when the QDs are antisymetrical and alligned parallel to each other.

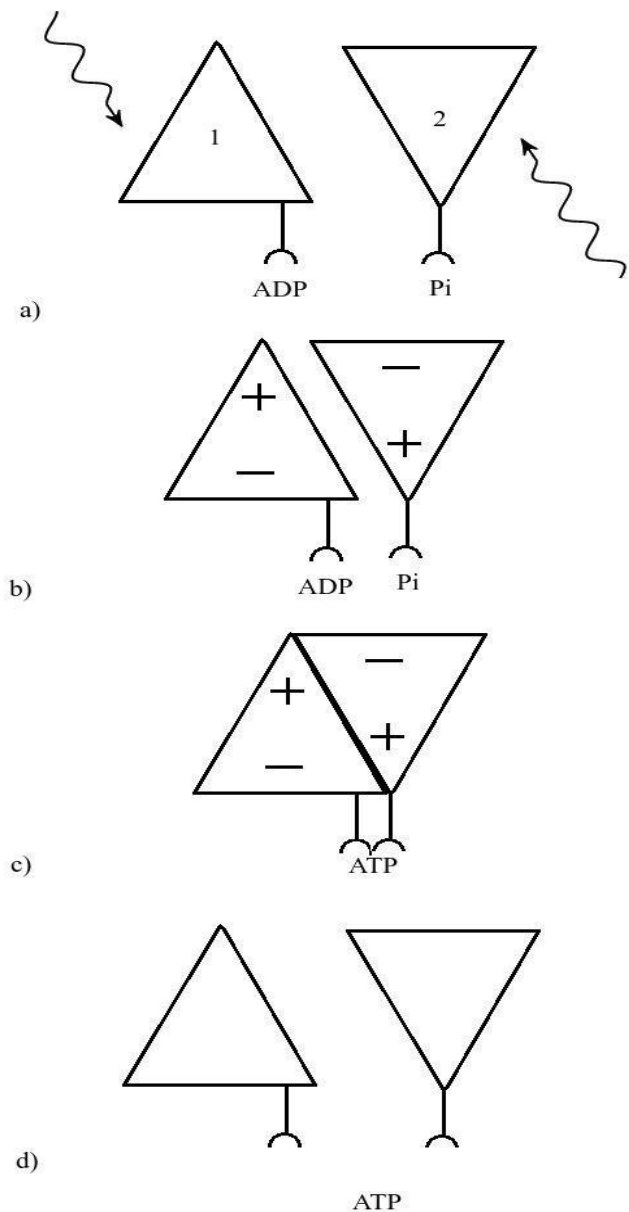

Fig. 2: Two quantum dots that are used to join together ADP and Pi groups

By surrounding the dot with ZnS, another semiconductor, we can stop the release of the toxic $Cd^{2+}$ ions in the body. This material has a larger energy gap than CdSe therefore we should be able to increase the lifetime of excitons (i.e. the time between the generation and the recombination of the electron-hole pair). We also mention that the quantum dot's structure can host more than one exciton. For simplicity we will refer to the core (shell) QD CdSe(Zns) that is encapsulated in cystein as QD. If an adenozine-diphosphate(ADP) molecule binds to the apex of a QD and a phosphate group binds on the margin of the base of the other QD, this would catalyze the reaction ADP+Pi=ATP(adenozine-triphosphate) through collisions between the two charged dots (Fig. 2 c).

After bringing together the ADP molecule and the Pi group by the Coulomb attraction of opposite charges we form the ATP molecule that should be able to be freed into the solution due to the energetic surplus.

The retinal exclusion threshold for human is 76.5±1.5 kDa (6.11±0.04 nm)[8] and the threshold fot the blood-brain-barrier is 0.5 kDa[9] therefore the QDs must have dimensions between these two limits.

At the dissociation of ATP into ADP and Pi 30kJ/mol are liberated which is the equivalent of 312 meV per QD. The energy gap of CdSe is 1740 meV (167 kJ/mol) therefore a single photon with the resonant energy has enough energy to combine

ADP and Pi into ATP. If we take into consideration the whole system the energy that is left over from the two photons is 3168 meV. This energy must be larger than the binding energy between ADP and dot 1 to which we add the binding energy between Pi and dot 2.

The decay time of an exciton in some nano-structures was shown to be as large as a few microseconds[10]. This means that the exciton is sufficiently persistent in order for the merging of the dots to occur at this scale of time. The computation of the minimum persistence time of an exciton can be done using the expression of the Coulomb potential between two charges found in a medium whose Debye-Hückel radius is of the order of that of the water. (i.e. 8 Å). This potential reads:

$$V(r) = -\frac{e^2}{4\pi\varepsilon_0 \varepsilon_r r} e^{\frac{-r}{D}},$$

where V(r) is the attraction potential, $\varepsilon_0$ is the vacuum dielectric permittivity, $\varepsilon_{eff}$ is the relative permittivity of water at room temperature (i.e. 78.33), D is the Debye-Hückel radius.

The mass of a dot is approximately m=60 kDa. By solving:

$$ma(t) = -\frac{dV(r)}{dr},$$

we find that the time between two consecutive bumps is τ=10 μs., Here a(t) is the acceleration of a dot towards their common mass center. In solving the differential equation the condition r(τ)=0 has to be set.

I mention that the ADP must get close enough to Pi in order for the reaction to occur. This distance is about 2 Å which is the breaking point of ATP [11]. Therefore the collision of the two QDs must bring the ADP and Pi within a distance of maximum 0.2 nm.

In [12] it has been reported the creation of colloidal solutions of bipyramidal core/shell

QDs of about 20nm made of CdSe/CdS.

There remains only on further step: conjugating the dots with molecules that dock to ADP and Pi. For this, the initial solution of QDs must be split in two separate solutions. The first one will be conjugated to streptavidin and a biotinylated low affinity adenosine receptor A2B-R (ADORA2B). The affinity of A2B receptor is approximately Ki =56nM[13]. This affinity corresponds to a variation of Gibbs energy $\Delta G = RT \ln K_i$ yielding a value of 41.37 kJ/mol.

It can be shown that the electrostatic force between two dipoles is greater for parallel dipoles lying in different planes than the force between two collinear dipoles at the same distance apart. This means that the configuration of dipoles depicted in Fig. 2 is favoured over the case in which the two dipoles are parallel and collinear.

There are many proteins that have binding sites for Pi but we are interested only in the ones that have low affinity and low molecular mass.

Two QDs that happen to be in vicinity will attract bringing thus together the ADP and phosphate molecules and making possible the reaction (see Fig. 2 b).

Once they come in contact the two dots are allowed to recombine their charge carriers through tunneling between dots. In this way the hole of the dot recombines with the electron from the base of the dot in the right and vice-versa. During their contact ATP is formed (Fig. 2c).

After charge neutralization the two dots will separate due to Brownian motion and this will release the ATP (Fig. 2d).

The QDs may be pyramidal or conical CdSe/ZnS dots. A hydrophilic polymer coating is prefferable. One ligand should be the Anti-Adenosine Receptor A2a antibody which has to be biotinylated and linked with streptavidin conjugated dots. The other ligand can be a PhosTag molecule conjugated in the same fashion as the previous molecule.

# ATP synthesis using fluorescence resonance energy transfer between two linked quantum dots

In this section I suggest another possible solution of generating ATP from ADP and Pi using pairs of linked QDs. Fluorescence Resonance Energy Transfer(FRET) technique was invented over 50 years ago and was since used for the transfer of vibrational energy between a donor and an acceptor molecule. The donor molecule can be a chromophore that absorbs the energy while the acceptor is the chromophore to which the energy is transferred. This phenomena are extensively used in biomedicine. The advantage over the previous method is that the energy transfer within this technique can be done without the collision of the acceptor and the donor molecules. Two molecules that interact in a way that FRET occurs form a donor/acceptor pair. The donor and acceptor must be close to each other (around 10-100 Å). The condition in order for FRET to occur between the molecules is that the spectrum of the donor molecule to overlap the energy spectrum of the acceptor molecule. Instead of molecules we can use QDs with the right energy gap.

QDs were shown to have been used as donors and acceptors but the idea that this transfer of energy could be used in biomedicine for treating RP was not suggested. The idea is the following: the energy of the photons that come in contact with the donor quantum dot is transmitted through resonance to the acceptor QD which is conjugated with anti-ADP (or anti-phosphate e.g. PhosTag) ligand molecules. Each phosphate gourp coming close to the energetic ADP conjugated to these ligands will react and form ATP. The energy that is necessary in order for this reaction to occur comes from the donor quantum dot. The transfer of energy between the donor and the acceptor dot is necessary because the frequency of light has to be converted to he frequency that corresponds to the ATP dissociation energy. This energy is small and the corresponding wavelength does not lie in the visible spectrum(E= 312 meV), therefore an energy conversion must occur. This conversion is guaranteed by the FRET of the two QDs.

Each of the dots has also conjugated a sequence of oligonucleotide, and the two oligonucleotides are complementary in order for the FRET binding to occur. Once bound the energy transfer via FRET is allowed.

I mention that the two QDs must have overlapping spectra and that they replace the mitochondrial organelle that was destroyed in the process of necrosis of the PCs. The donor QD and the oligonucleotide must have a total molecular weight that is smaller than the  maximum allowed molecular weight of the BRB. The acceptor dot and the conjugated streptavidin with biotinylized anti-ADP must also have an overall molecular of less than 76 kDa in order to pass through the BRB. One has to find right sizes of the dots in order for the energy transfer between the two dots to occur.

**Apoptosis reversal using quantum dots**

In this section it is suggested a method that should be able to bring back together the organelles of a broken PC that underwent apoptosis.

In apoptotic regulated cell death organelles are encapsulated within their own membrane. Being separated from a functional point of view from other organelles the cell's function is therefore lost. Grouping them back piece by piece can be a daunting task yet it may happen from an energetic point of view because the energy contribution of the surface tension of individually encapsulated organelles is higher than the surface tension energy of the healthy intact cell. By linking the organelles together using a nanowire the organelles can be brought back together. The organelles are to be found at  6 microns apart after one minute of apoptosis according to the Brownian law of motion. If the nanowires have radii of about 15 nanometers then their height has to be about 2 micrometers in order for the water in the cytoplasm of an organelle to climb up along the dot and connect with the other organelle. In order to make the coalescing of organelles more feasible the dot that was introduced earlier has to be broken down in two parts, each one of about 1 micrometer in length and each one of them having to be able to anchor itself on the membrane of one part of the broken cell to be further joined. If the dot is further made of magnetite which is a magnetic biodegradable material the two parts should attract each other and unite into a bridge (see Fig. 3c)

In conclusion a nanowire made of magnetite should be able to conjugate its end to the membrane of the cell's organelles that underwent apoptosis. Two such nanowires should attract each other by the other free magnetic end. Two nanowires should have opposites poles in order for the coalescing to occur. Therefore the place of the anchored ligands are inverted.

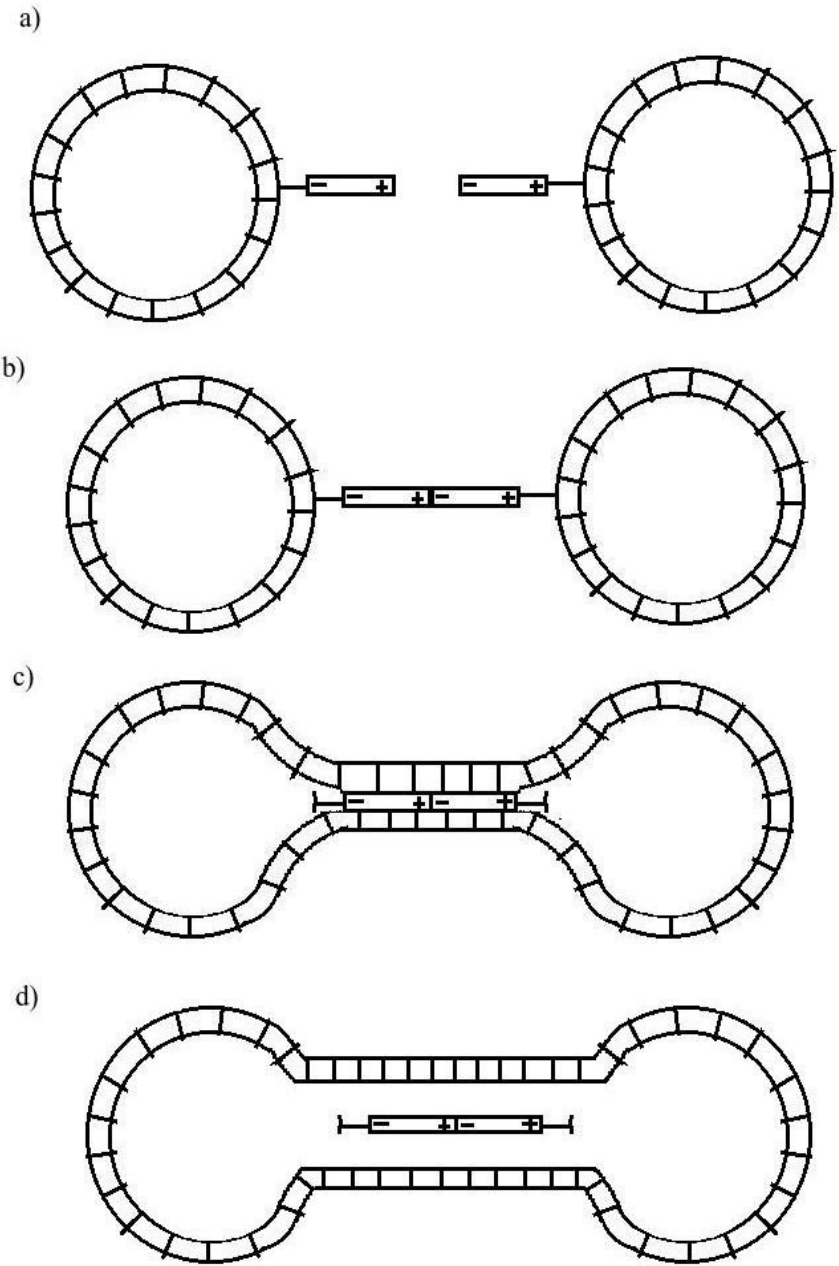

*Fig. 3: Two magnetite quantum dots used to join together organelles after apoptosis*

**References**

[1] Maria A. Musarella and Ian M. MacDonald, Current Concepts in the Treatment of Retinitis Pigmentosa, Journal of Ophtalmology, 18(2), 2011, 112-116. (8)

[2] Y Murakami, Y Ikeda, S Nakatake, J W Miller, D G Vavvas, K H Sonoda and T Ishibashi, Necrotic cone photoreceptor cell death in retinitis pigmentosa, Cell Death and Disease, 6, 2015, e2038,

[3] Derfus A M, Chan W C W and Bhatia, Probing the cytotoxicity of semiconductor quantum dots Nano Lett. 4, 2004, 11-8,

[4] Shu-quan Chang, Bin Kang, Yao-dong Dai, Hong-xu Zhang and Da Chen, One-step fabrication of biocompatible chitosan-Coated ZnS and ZnS:$Mn^{2+}$ quantum Dots via a Γ-Radiation Route, Nanoscale Res Lett, 6(1), 2011, 591

[5] Shichen Lian, David J. Weinberg, Rachel D. Harris, Mohamad S. Kodaimati, and Emily A. Weiss, Subpicosecond Photoinduced Hole Transfer from a CdS Quantum Dot to a Molecular Acceptor Bound Through an Exciton-Delocalizing Ligand, ACS Nano,10 (6), 2016, 6372–6382

[6] Nianhui Song, Haiming Zhu, Shengye Jin, and Tianquan Lian, Hole Transfer from Single Qauntum Dots, ACS Nano, 5(11), 2011, 8750-8759

[7] Dubertret B, Skourides P, Norris DJ, Noireaux V, Brivanlou AH, Libchaber A. In vivo imaging of quantum dots encapsulated in phospholipid micelles, Science 298(5599), 2002, 1759-62.

[8] Timothy L. Jackson; Richard J. Antcliff; Jost Hillenkamp; John Marshall, Human Retinal Molecular Weight Exclusion Limit and Estimate of Species Variation,

Investigative Ophthalmology & Visual Science, Vol.44, 2141-2146 (2003)

[9] William M. Pardridge NeuroRx,The Blood-Brain Barrier: Bottleneck in Brain Drug Development, NeuroRX, 2(1): 3–14. (2005)

[10] Beeler, M. and Lim, C. B. and Hille, P. and Bleuse, J. and Schörmann, J. and de la Mata, M. and Arbiol, J. and Eickhoff, M. and Monroy, E., Long-lived excitons in GaN/AlN nanowire heterostructures, Phys. Rev. B , 91,20,205440,8 (2015)

[11] Alan E. Senior, Sasha Nadanaciva, Joachim Weber, The molecular mechanism of ATP synthesis by F1F0-ATP synthase, Biochimica et Biophysica Acta,1553, 188-211 (2002)

[12] Arjen Toni Dijksman, Spectroscopy of Colloidal Quantum Dots of Controlled Shape and Size. Micro and nanotechnologies/Microelectronics. Université Pierre et Marie Curie – Paris VI, 2013. English. <pastel-00876357>

[13] Constance N. Wilson, S. Jamal Mustafa, Adenosine Receptors in Health and Disease, Springer, page 103 (2009)